\documentclass[12pt,titlepage]{article}
\usepackage{amsfonts}
\usepackage[portuguese]{babel}

\begin{document}

\title{\textbf{Sobre o Limiar para a Produ\c{c}\~{a}o de Pares e Localiza%
\c{c}\~{a}o de Part\'{\i}culas sem Spin}\\
{\small On the Threshold for the Pair Production and Localization of \
Spinless Particles}}
\author{Tatiana R. Cardoso e Antonio S. de Castro\thanks{
E-mail: castro@pesquisador.cnpq.br.} \\
\\
UNESP - Campus de Guaratinguet\'{a}\\
Departamento de F\'{\i}sica e Qu\'{\i}mica\\
12516-410 Guaratinguet\'{a} SP - Brasil}
\date{}
\date{}
\maketitle

\begin{abstract}
\noindent A equa\c{c}\~{a}o de Klein-Gordon em uma dimens\~{a}o espacial
\'{e} investigada com a mais geral estrutura de Lorentz para os potenciais
externos. A an\'{a}lise do espalhamento de part\'{\i}culas em um potencial
degrau com uma mistura arbitr\'{a}ria de acoplamentos vetorial e escalar
revela que o acoplamento escalar contribui para aumentar o limiar da energia
de produ\c{c}\~{a}o de pares. Mostra-se ainda que a produ\c{c}\~{a}o de
pares torna-se fact\'{\i}vel somente quando o acoplamento vetorial excede o
acoplamento escalar. Um aparente paradoxo relacionado com a localiza\c{c}%
\~{a}o de uma part\'{\i}cula em uma regi\~{a}o do espa\c{c}o arbitrariamente
pequena, devido \`{a} presen\c{c}a do potencial escalar, \'{e}
re\-sol\-vi\-do com a introdu\c{c}\~{a}o do conceito de comprimento de onda
Compton efetivo. \newline
\newline
\noindent \textbf{Palavras-chave:} equa\c{c}\~{a}o de Klein-Gordon, paradoxo
de Klein, produ\c{c}\~{a}o de pares, localiza\c{c}\~{a}o, comprimento de
onda Compton \newline
\newline
\newline
\noindent {\small {The one-dimensional Klein-Gordon equation is investigated
with the most general Lorentz structure for the external potentials. The
analysis of the scattering of particles in a step potential with an
arbitrary mixing of vector and scalar couplings reveals that the scalar
coupling contributes for increasing the threshold energy for the
particle-antiparticle pair production. Furthermore, it is shown that the
pair production is only feasible whether the vector coupling exceeds the
scalar one. An apparent paradox concerning the localization of a particle in
an arbitrarily small region of space, due to the presence of the scalar
coupling, is solved by introducing the concept of effective Compton
wavelength.\newline
\newline
\noindent \textbf{Keywords:} Klein-Gordon equation, Klein\'{}s paradox, pair
production, localization, Compton wavelength} }
\end{abstract}

\section{Introdu\c{c}\~{a}o}

A generaliza\c{c}\~{a}o da mec\^{a}nica qu\^{a}ntica que inclui a
relatividade especial \'{e} ne\-ces\-s\'{a}\-ria para a descri\c{c}\~{a}o de
fen\^{o}menos em altas energias e tamb\'{e}m para a descri\c{c}\~{a}o de fen%
\^{o}menos em escalas de comprimentos que s\~{a}o menores ou compar\'{a}veis
com o comprimento de onda Compton da part\'{\i}cula ($\lambda=\hbar /(mc)
$). A generaliza\c{c}\~{a}o n\~{a}o \'{e} uma tarefa trivial e novos e
peculiares fen\^{o}menos surgem na Mec\^{a}nica Qu\^{a}ntica Relativ\'{\i}%
stica (doravante denominada MQR). Entre tais fen\^{o}menos est\~{a}o a produ%
\c{c}\~{a}o espont\^{a}nea de pares mat\'{e}ria-antimat\'{e}ria e a limita%
\c{c}\~{a}o para a lo\-ca\-li\-za\-\c{c}\~{a}o de part\'{\i}culas. Essa
limita\c{c}\~{a}o pode ser estimada \ pela observa\c{c}\~{a}o que a m\'{a}%
xima incerteza para o momento da part\'{\i}cula $\Delta p=mc$ conduz, via
princ\'{\i}pio da incerteza de Heisenberg, \`{a} incerteza m\'{\i}nima na
posi\c{c}\~{a}o $\Delta x=\lambda/2$ \cite{gre}-\cite{str}. Embora a MQR
como modelo de part\'{\i}cula \'{u}nica, referida como formalismo de
primeira quantiza\c{c}\~{a}o, n\~{a}o possa dar conta da completa descri\c{c}%
\~{a}o da cria\c{c}\~{a}o de pares, ela pavimenta o caminho para o
desenvolvimento da Teoria Qu\^{a}ntica de Campos.

As mais simples equa\c{c}\~{o}es da MQR s\~{a}o a equa\c{c}\~{a}o de
Klein-Gordon (EKG)\footnote{%
A EKG descreve o comportamento de b\'{o}sons de spin 0. P\'{\i}ons e k\'{a}%
ons, por exemplo.} e a equa\c{c}\~{a}o de Dirac\footnote{%
A equa\c{c}\~{a}o de Dirac descreve o comportamento de f\'{e}rmions de spin
1/2, tais como o el\'{e}tron, o neutrino, o quark, o pr\'{o}ton e o n\^{e}%
utron.
\par
{}}. O spin \'{e} uma complica\c{c}\~{a}o adicional na MQR e, naturalmente,
a EKG permite que certos aspectos da MQR possam ser analisados com um
formalismo matem\'{a}tico mais simples e percebidos com maior transpar\^{e}%
ncia. A solu\c{c}\~{a}o da equa\c{c}\~{a}o de Dirac para o espalhamento de
part\'{\i}culas em um potencial degrau, considerado como o componente
temporal de um potencial vetorial, \'{e} bem conhecida e cristalizada em
livros-texto \cite{gre}-\cite{gro}. Neste problema surge o c\'{e}lebre
paradoxo de Klein \cite{kle} para potenciais suficientemente intensos, um fen%
\^{o}meno em que o coeficiente de reflex\~{a}o excede a unidade e \'{e}
interpretado como sendo devido \`{a} cria\c{c}\~{a}o de pares na interface
do potencial. A an\'{a}lise do problema consoante a EKG n\~{a}o foi
esquecida \cite{gro}, \cite{wi}-\cite{vil}.

Neste trabalho analisamos a EKG unidimensional com intera\c{c}\~{o}es
externas com a mais geral estrutura de Lorentz, i. e., consideramos
potenciais com estrutura vetorial, com componentes espacial e temporal,
acrescido de uma estrutura escalar. Em seguida exploramos as solu\c{c}\~{o}%
es para o espalhamento de part\'{\i}culas em um potencial degrau com
acoplamento geral, por assim dizer, com uma mistura arbitr\'{a}ria de
acoplamentos vetorial e escalar. Verificamos que tal mistura de acoplamentos
conduz a resultados surpreendentes. Para al\'{e}m de aumentar o limiar de
energia para a produ\c{c}\~{a}o espont\^{a}nea de pares, podendo at\'{e}
mesmo frustrar a produ\c{c}\~{a}o ainda que os potenciais sejam extremamente
fortes, a presen\c{c}a de um acoplamento escalar permite que uma part\'{\i}%
cula possa ser localizada em uma regi\~{a}o do espa\c{c}o arbitrariamente
pequena sem amea\c{c}ar a interpreta\c{c}\~{a}o de part\'{\i}cula \'{u}nica
da EKG. A aparente viola\c{c}\~{a}o do princ\'{\i}pio da incerteza \'{e}
remediada com a introdu\c{c}\~{a}o do conceito de comprimento de onda
Compton efetivo.

Apesar da originalidade e generalidade, este trabalho \'{e} acess\'{\i}vel
aos estudantes de gradua\c{c}\~{a}o em f\'{\i}sica que tenham freq\"{u}%
entado alguns poucos meses de um curso introdut\'{o}rio de mec\^{a}nica qu%
\^{a}ntica. Dessa forma permite-se o acesso precoce de estudantes a alguns
dos mais interessantes fen\^{o}menos da MQR.

\section{A equa\c{c}\~{a}o de Klein-Gordon}

A EKG unidimensional para uma part\'{\i}cula livre de massa de repouso $m$
cor\-res\-pon\-de \`{a} rela\c{c}\~{a}o energia-momento relativ\'{\i}stica $%
E^{2}=c^{2}p^{2}+m^{2}c^{4}$, onde a energia $E$ e o momento $p$ tornam-se
operadores, $i\hbar \,\partial /\partial t$ e $-i\hbar \,\partial /\partial
x $ respectivamente, atuando sobre a fun\c{c}\~{a}o de onda $\Phi (x,t)$.
Aqui, $c$ \'{e} a velocidade da luz e $\hbar $ \'{e} a constante de Planck ($%
\hbar =h/(2\pi )$).

Na presen\c{c}a de potenciais externos a rela\c{c}\~{a}o energia-momento
torna-se%
\begin{equation}
\left( E-V_{t}\right) ^{2}=c^{2}\left( p-\frac{V_{e}}{c}\right) ^{2}+\left(
mc^{2}+V_{s}\right) ^{2}  \label{1}
\end{equation}%
onde os subscritos nos termos dos potenciais denotam suas propriedades com
res\-pei\-to \`{a}s transforma\c{c}\~{o}es de Lorentz: $t$ e $e$ para os
componentes temporal e espacial de um potencial vetorial\footnote{%
A energia e o momento s\~{a}o os componentes temporal e espacial,
respectivamente, da quantidade $(E/c\,,\,p)$, a qual se comporta, segundo as
transforma\c{c}\~{o}es de Lorentz, como um vetor. O potencial vetorial, com
componentes $(V_{t}\,,\,V_{e})$, \'{e} acoplado \`{a} part\'{\i}cula de
acordo com o \textit{princ\'{\i}pio do acoplamento m\'{\i}nimo}, tamb\'{e}m
chamado de \textit{princ\'{\i}pio da substitui\c{c}\~{a}o m\'{\i}nima}, $%
E\rightarrow E-V_{t}$ e $p\rightarrow p-V_{e}/c$, como \'{e} habitual no
caso da intera\c{c}\~{a}o eletromagn\'{e}tica.}, e $s$ para um potencial
escalar\footnote{%
A massa de repouso \'{e} uma quantidade invariante de Lorentz, i. e., uma
quantidade escalar. O potencial escalar foi acoplado \`{a} part\'{\i}cula em
(1) de acordo com o \textit{princ\'{\i}pio do acoplamento m\'{\i}nimo} $%
m\rightarrow m+V_{s}/c^{2}$. Esta prescri\c{c}\~{a}o fornece o limite n\~{a}%
o-relativ\'{\i}stico apropriado da EKG, conforme veremos adiante, em
contraste com a regra $m^{2}\rightarrow m^{2}+V_{s}^{2}/c^{4}$ empregada na
Ref. \cite{gre}.}.

A equa\c{c}\~{a}o da continuidade para a EKG%
\begin{equation}
\frac{\partial \rho }{\partial t}+\frac{\partial J}{\partial x}=0
\label{con}
\end{equation}
\noindent \'{e} satisfeita com $\rho $ e $J$ definidos como%
\begin{eqnarray}
\rho &=&\frac{i\hbar }{2mc^{2}}\left( \Phi ^{\ast }\frac{\partial \Phi }{%
\partial t}-\frac{\partial \Phi ^{\ast }}{\partial t}\Phi \right) -\frac{%
V_{t}}{mc^{2}}\left\vert \Phi \right\vert ^{2}  \nonumber \\
&&  \label{2} \\
J &=&\frac{\hbar }{2im}\left( \Phi ^{\ast }\frac{\partial \Phi }{\partial x}-%
\frac{\partial \Phi ^{\ast }}{\partial x}\Phi \right) -\frac{V_{e}}{mc}%
\left\vert \Phi \right\vert ^{2}  \nonumber
\end{eqnarray}%
Vale a pena observar o modo que os componentes do potencial vetorial
participam da densidade $\rho $ e da corrente $J$, tanto quanto a aus\^{e}%
ncia do potencial escalar. Observa-se tamb\'{e}m que a densidade envolve
derivadas temporais, um fato relacionado com a derivada temporal de segunda
ordem na EKG, e pode admitir valores negativos mesmo no caso de uma part%
\'{\i}cula livre. Assim sendo $\rho $ n\~{a}o pode ser interpretada como uma
densidade de probabilidade. Contudo, Pauli e Weisskopf \cite{pau} mostraram
que n\~{a}o h\'{a} dificuldade com a interpreta\c{c}\~{a}o da densidade e da
corrente da EKG se essas grandezas forem interpretadas como densidade e
corrente de \textit{carga}, ao inv\'{e}s de densidade e corrente de
probabilidade. A \textit{carga}\ n\~{a}o deve ser pensada necessariamente
como carga el\'{e}trica, mas como carga generalizada que satisfaz uma lei de
conserva\c{c}\~{a}o aditiva, por assim dizer que a \textit{carga} de um
sistema \'{e} a soma das \textit{cargas} de suas partes constituintes.

Para potenciais externos independentes do tempo, a EKG admite solu\c{c}\~{o}%
es da forma%
\begin{equation}
\Phi (x,t)=\phi (x)\,e^{i\Lambda \left( x\right) }\,e^{-i\frac{E}{\hbar }t}
\label{2a}
\end{equation}

\noindent onde $\phi $ obedece a uma equa\c{c}\~{a}o similar em forma \`{a}
equa\c{c}\~{a}o de Schr\"{o}dinger

\begin{equation}
-\frac{\hbar ^{2}}{2m}\,\frac{d^{2}\phi }{dx^{2}}+\left( \frac{%
V_{s}^{2}-V_{t}^{2}}{2mc^{2}}+V_{s}+\frac{E}{mc^{2}}\,V_{t}\right) \,\phi =%
\frac{E^{2}-m^{2}c^{4}}{2mc^{2}}\,\phi  \label{3}
\end{equation}%
com $\Lambda \left( x\right) =\int^{x}dy\,V_{e}(y)/(\hbar c)$. A elimina\c{c}%
\~{a}o do componente espacial do potencial vetorial \'{e} equivalente a uma
redefini\c{c}\~{a}o do operador momento. Realmente,
\begin{equation}
\left( p_{op}-\frac{V_{e}}{c}\right) ^{2}\Phi =e^{i\Lambda
}\,p_{op}^{2}\,\phi  \label{mom}
\end{equation}

\'{E} agora importante perceber que h\'{a} solu\c{c}\~{o}es de energia
positiva tanto quanto solu\c{c}\~{o}es de energia negativa\footnote{%
As solu\c{c}\~{o}es de energia positiva e negativa s\~{a}o assossiadas com
part\'{\i}culas e antipart\'{\i}culas, respectivamente.} e que os dois poss%
\'{\i}veis sinais para $E$ implicam em duas possibilidades para a evolu\c{c}%
\~{a}o temporal da fun\c{c}\~{a}o de onda. Seja como for, a energia \'{e}
uma quantidade conservada. A forma da equa\c{c}\~{a}o de autovalor (\ref{3})
\'{e} preservada sob as transforma\c{c}\~{o}es simult\^{a}neas $E\rightarrow
-E$ e $V_{t}\rightarrow -V_{t}$, e isto implica que part\'{\i}culas e
antipart\'{\i}culas est\~{a}o sujeitas a componentes temporais de um
potencial vetorial com sinais dissimilares. Como conseq\"{u}\^{e}ncia
imediata dessa covari\^{a}ncia tem-se que, \ por mais estranho que possa
parecer, part\'{\i}culas e antipart\'{\i}culas compartilham exatamente a
mesma autofun\c{c}\~{a}o no caso de um potencial puramente escalar e que o
espectro \'{e} disposto simetricamente em torno de $E=0$. \textit{Cargas}
positivas e negativas est\~{a}o sujeitas a acoplamentos vetoriais
(componentes temporais) de sinais contr\'{a}rios e igual acoplamento
escalar. A intera\c{c}\~{a}o escalar \'{e} independente da \textit{carga} e
assim age indiscriminadamente sobre part\'{\i}culas e antipart\'{\i}culas.
Diz-se ent\~{a}o que o potencial vetorial acopla com a \textit{carga} da part%
\'{\i}cula e que o potencial escalar acopla com a massa da part\'{\i}cula. \
A densidade e a corrente correspondentes \`{a} solu\c{c}\~{a}o expressa por (%
\ref{2a}) tornam-se%
\begin{eqnarray}
\rho &=&\frac{E-V_{t}}{mc^{2}}\left\vert \phi \right\vert ^{2}  \nonumber \\
&&  \label{3a} \\
J &=&\frac{\hbar }{2im}\left( \phi ^{\ast }\frac{\partial \phi }{\partial x}-%
\frac{\partial \phi ^{\ast }}{\partial x}\phi \right)  \nonumber
\end{eqnarray}%
Em virtude de $\rho $ e $J$ serem independentes do tempo, a solu\c{c}\~{a}o (%
\ref{2a}) \'{e} dita descrever um estado estacion\'{a}rio. Nota-se que a
densidade torna-se negativa em regi\~{o}es do espa\c{c}o onde $V_{t}>E$ e
que o componente espacial do potencial vetorial n\~{a}o mais interv\'{e}m na
corrente.

Ademais, deve-se mencionar que a EKG reduz-se \`{a} \ equa\c{c}\~{a}o de Schr%
\"{o}dinger no limite n\~{a}o-relativ\'{\i}stico ($E\simeq mc^{2}$ e
energias potenciais pequenas comparadas com $mc^{2}$) com $\phi $ obedecendo
\ \`{a} equa\c{c}\~{a}o%
\begin{equation}
-\frac{\hbar ^{2}}{2m}\,\frac{d^{2}\phi }{dx^{2}}+\left( V_{t}+V_{s}\right)
\phi =\left( E-mc^{2}\right) \phi  \label{3b}
\end{equation}%
No limite n\~{a}o-relativ\'{\i}stico as naturezas de Lorentz dos potenciais n%
\~{a}o sofrem quaisquer distin\c{c}\~{o}es, e a densidade e a corrente
reduzem-se exatamente aos va\-lo\-res da teoria n\~{a}o-relativ\'{\i}stica.

\section{A solu\c{c}\~{a}o para um potencial degrau}

Vamos agora considerar a EKG com os potenciais externos independentes do
tempo na forma de um degrau de potencial. Consideramos $V_{e}=0$, haja vista
que o componente espacial do potencial vetorial contribui apenas com um
fator de fase local para $\Phi (x,t)$ e n\~{a}o contribui para a densidade
nem para a corrente. O potencial degrau \'{e} expresso como
\begin{equation}
V(x)=\left\{
\begin{array}{c}
0\quad \mathrm{para}\quad x<0 \\
V_{0}\quad \mathrm{para}\quad x>0%
\end{array}%
\right.  \label{4}
\end{equation}%
onde $V_{0}>0$. Os potenciais vetorial e escalar s\~{a}o escritos como $%
V_{t}(x)=g_{t}V(x)$ e $V_{s}(x)=g_{s}V(x)$ de tal forma que as constantes de
acoplamento est\~{a}o sujeitas ao v\'{\i}nculo $g_{t}+g_{s}=1$, com $%
g_{t}\geq 0$ e $g_{s}\geq 0$.

Para $x<0$, a EKG apresenta solu\c{c}\~{o}es na forma de uma soma de autofun%
\c{c}\~{o}es do operador momento:%
\begin{equation}
\phi =A_{+}\,e^{+ikx}+A_{-}\,e^{-ikx}  \label{5}
\end{equation}%
onde
\begin{equation}
k=\frac{\sqrt{E^{2}-m^{2}c^{4}}}{\hbar c}  \label{6}
\end{equation}%
Para $|E|>mc^{2}$, a solu\c{c}\~{a}o expressa por (\ref{5}) reverte-se em
ondas planas pro\-pa\-gan\-do-se em ambos os sentidos do eixo $X$ com
velocidade de grupo\footnote{%
Veja, e.g., Refs. [1] e [5].}%
\begin{equation}
v_{g}=\frac{1}{\hbar }\,\frac{dE}{dk}  \label{7}
\end{equation}%
igual \`{a} velocidade cl\'{a}ssica da part\'{\i}cula. Se escolhermos part%
\'{\i}culas incidindo sobre a barreira de potencial ($E>mc^{2}$) teremos que
$A_{+}\,e^{+ikx}$ descreve part\'{\i}culas incidentes ($v_{g}=c^{2}\hbar
k/E>0$), enquanto $A_{-}\,e^{-ikx}$ descreve part\'{\i}culas refletidas ($%
v_{g}=-c^{2}\hbar k/E<0$). $\ $ A corrente nesta regi\~{a}o do espa\c{c}o,
correspondendo a $\phi $ dada por \ (\ref{5}), \'{e} expressa por
\begin{equation}
J=\frac{\hbar k}{m}\left( |A_{+}|^{2}-|A_{-}|^{2}\right)  \label{7a}
\end{equation}%
Observe que a rela\c{c}\~{a}o $J=\rho \,v_{g}$ mant\'{e}m-se tanto para a
onda incidente quanto para a onda refletida pois%
\begin{equation}
\rho =\frac{E}{mc^{2}}\,|\phi |^{2}>0  \label{7b}
\end{equation}

Por outro lado, para $x>0$ devemos ter $v_{g}\geq 0$ de forma que a solu\c{c}%
\~{a}o nesta regi\~{a}o do espa\c{c}o descreve uma onda evanescente ou uma
onda progressiva que se afasta da interface do potencial. A solu\c{c}\~{a}o
geral tem a forma%
\begin{equation}
\phi =B_{+}\,e^{+i\kappa x}+B_{-}\,e^{-i\kappa x}  \label{8}
\end{equation}%
onde
\begin{equation}
\kappa =\frac{\sqrt{\left( E-g_{t}V_{0}\right) ^{2}-\left(
mc^{2}+g_{s}V_{0}\right) ^{2}}}{\hbar c}  \label{9}
\end{equation}%
Por causa da dupla possibilidade de sinais para a energia de um estado
estacion\'{a}rio, a solu\c{c}\~{a}o $B_{-}\,e^{-i\kappa x}$ n\~{a}o pode ser
descartada a priori. De fato, pode-se depreender de (\ref{2a}) que esta
parcela pode vir a descrever uma onda progressiva com energia negativa e
velocidade de fase $v_{f}=|E|/(\hbar \kappa )>0$. Percebe-se claramente que
podemos segregar tr\^{e}s classes distintas de solu\c{c}\~{o}es:

\medskip

\begin{itemize}
\item \textbf{Classe A. } Para $V_{0}<E-mc^{2}$ temos que $\kappa \in
\mathbb{R}
$ e a solu\c{c}\~{a}o que descreve ondas planas propagando-se no sentido
positivo do eixo $X$ com velocidade de grupo%
\begin{equation}
v_{g}=\frac{c^{2}\hbar \kappa }{E-g_{t}V_{0}}  \label{10}
\end{equation}%
\'{e} poss\'{\i}vel somente se $B_{-}=0$. Neste caso, a densidade e a
corrente s\~{a}o dadas por%
\begin{equation}
\rho =\frac{E-g_{t}V_{0}}{mc^{2}}\,|B_{+}|^{2}\quad \mathrm{e}\quad J=\frac{%
\hbar \kappa }{m}\,|B_{+}|^{2}  \label{10a}
\end{equation}

\item \textbf{Classe B. } Para $E-mc^{2}<V_{0}<$ $V_{c}$, onde%
\begin{equation}
V_{c}=\left\{
\begin{array}{c}
\frac{E+mc^{2}}{2g_{t}-1} \\
\\
\infty%
\end{array}%
\begin{array}{c}
{\textrm{para }}g_{t}>\frac{1}{2} \\
\\
{\textrm{para }}g_{t}\leq \frac{1}{2}%
\end{array}%
\right.  \label{10aa}
\end{equation}
temos que $\kappa =i|\kappa |$ de forma que (\ref{8}), com $B_{-}=0$%
\footnote{%
A condi\c{c}\~{a}o $B_{-}=0$ \'{e} necess\'{a}ria para que a densidade seja
finita quando $x\rightarrow +\infty $.}, descreve uma onda evanescente.
Neste caso,%
\begin{equation}
\rho =\frac{E-g_{t}V_{0}}{mc^{2}}\,|B_{+}|^{2}\,e^{-2|\kappa |x}\quad
\mathrm{e}\quad J=0  \label{10b}
\end{equation}
\end{itemize}

\medskip

\begin{itemize}
\item \textbf{Classe C. } $V_{0}>V_{c}$, com $V_{c}$ concebido na classe
\textbf{B}, surge mais uma vez a possibilidade de propaga\c{c}\~{a}o no
sentido positivo do eixo $X$, desta feita com $B_{+}=0$, com velocidade de
grupo%
\begin{equation}
v_{g}=\frac{c^{2}\hbar \kappa }{g_{t}V_{0}-E}  \label{11}
\end{equation}%
Nesta circunst\^{a}ncia em que o acoplamento vetorial excede o acoplamento
escalar nos defrontamos com um caso bizarro, pois tanto a densidade quanto a
corrente s\~{a}o quantidades negativas, viz.%
\begin{equation}
\rho =\frac{E-g_{t}V_{0}}{mc^{2}}\,|B_{-}|^{2}\quad \mathrm{e}\quad J=-\frac{%
\hbar \kappa }{m}\,|B_{-}|^{2}  \label{11a}
\end{equation}%
A manten\c{c}a da rela\c{c}\~{a}o $J=\rho \,v_{g}$, contudo, \'{e} uma licen%
\c{c}a para interpretar $B_{-}\,e^{-i\kappa x}$ a descrever a propaga\c{c}%
\~{a}o, no sentido positivo do eixo $X$, de part\'{\i}culas com \textit{carga%
}\ de sinal contr\'{a}rio ao das part\'{\i}culas incidentes. Esta interpreta%
\c{c}\~{a}o \'{e} consistente se as part\'{\i}culas pro\-pa\-gan\-do-se
nessa regi\~{a}o t\^{e}m energia $-E$ e est\~{a}o sob a influ\^{e}ncia de um
potencial vetorial $-g_{t}V_{0}$. Quer dizer, ent\~{a}o, que a onda
progressiva descreve, de fato, a propaga\c{c}\~{a}o de antipart\'{\i}culas
no sentido positivo do eixo $X$\footnote{%
Note que part\'{\i}cula e antipart\'{\i}cula t\^{e}m massas iguais.}.
\end{itemize}

\medskip

\subsection{\protect\medskip Os coeficientes de reflex\~{a}o e transmiss\~{a}%
o}

N\~{a}o obstante a descontinuidade do potencial em $x=0$, a autofun\c{c}\~{a}%
o e sua derivada primeira s\~{a}o fun\c{c}\~{o}es cont\'{\i}nuas\footnote{%
Esta conclus\~{a}o, v\'{a}lida para potenciais com descontinuidades finitas,
pode ser obtida pela integra\c{c}\~{a}o da Eq. (\ref{3}) entre $-\varepsilon
$ e $+\varepsilon $ no limite $\varepsilon \rightarrow 0$. Pode-se
verificar, pelo mesmo pro\-ce\-di\-men\-to, que apenas as autofun\c{c}\~{o}%
es s\~{a}o cont\'{\i}nuas quando as descontinuidades dos potenciais s\~{a}o
infinitas.}. A demanda por continuidade de $\phi $ e $d\phi /dx$ fixa as
amplitudes de onda em termos da amplitude da onda incidente $A_{+}$, viz.%
\begin{equation}
\frac{A_{-}}{A_{+}}=\left\{
\begin{array}{c}
\frac{k-\kappa }{k+\kappa } \\
\\
\frac{\left( k-i|\kappa |\right) ^{2}}{k^{2}+|\kappa |^{2}} \\
\\
\frac{k+\kappa }{k-\kappa }%
\end{array}%
\begin{array}{c}
{\textrm{para a classe }\mathbf{A}} \\
\\
{\textrm{para a classe }\mathbf{B}} \\
\\
{\textrm{para a classe }\mathbf{C}}%
\end{array}%
\right.  \label{12}
\end{equation}%
\begin{equation}
\frac{B_{+}}{A_{+}}=\left\{
\begin{array}{c}
\frac{2k}{k+\kappa } \\
\\
\frac{2k\left( k-i|\kappa |\right) }{k^{2}+|\kappa |^{2}} \\
\\
0%
\end{array}%
\begin{array}{c}
{\textrm{para a classe }\mathbf{A}} \\
\\
{\textrm{para a classe }\mathbf{B}} \\
\\
{\textrm{para a classe }\mathbf{C}}%
\end{array}%
\right.  \label{13}
\end{equation}%
\begin{equation}
\frac{B_{-}}{A_{+}}=\left\{
\begin{array}{c}
0 \\
\\
0 \\
\\
\frac{2k}{k-\kappa }%
\end{array}%
\begin{array}{c}
{\textrm{para a classe }\mathbf{A}} \\
\\
{\textrm{para a classe }\mathbf{B}} \\
\\
{\textrm{para a classe }\mathbf{C}}%
\end{array}%
\right.  \label{14}
\end{equation}%
\ \ \ Agora focalizamos nossa aten\c{c}\~{a}o na determina\c{c}\~{a}o dos
coeficientes de reflex\~{a}o $R$ e transmiss\~{a}o $T$. O coeficiente de
reflex\~{a}o (transmiss\~{a}o) \'{e} definido como a raz\~{a}o entre as
correntes refletida (transmitida) e incidente. Haja vista que $\partial \rho
/\partial t=0$ para estados estacion\'{a}rios, temos que a corrente \'{e}
independente de $x$. Usando este fato obtemos prontamente que%
\begin{equation}
R=\frac{|A_{-}|^{2}}{|A_{+}|^{2}}=\left\{
\begin{array}{c}
\left( \frac{k-\kappa }{k+\kappa }\right) ^{2} \\
\\
1 \\
\\
\left( \frac{k+\kappa }{k-\kappa }\right) ^{2}%
\end{array}%
\begin{array}{c}
{\textrm{para a classe }\mathbf{A}} \\
\\
{\textrm{para a classe }\mathbf{B}} \\
\\
{\textrm{para a classe }\mathbf{C}}%
\end{array}%
\right.  \label{15}
\end{equation}%
\begin{equation}
T=\left\{
\begin{array}{c}
\frac{\kappa }{k}\frac{|B_{+}|^{2}}{|A_{+}|^{2}}=\frac{4k\kappa }{\left(
k+\kappa \right) ^{2}} \\
\\
0 \\
\\
-\frac{\kappa }{k}\frac{|B_{-}|^{2}}{|A_{+}|^{2}}=-\frac{4k\kappa }{\left(
k-\kappa \right) ^{2}}%
\end{array}%
\begin{array}{c}
{\textrm{para a classe }\mathbf{A}} \\
\\
{\textrm{para a classe }\mathbf{B}} \\
\\
{\textrm{para a classe }\mathbf{C}}%
\end{array}%
\right.  \label{16}
\end{equation}

Em todas as classes temos que $R+T=1$, como deve ser. Entretanto, a classe
\textbf{C} apresenta $R>1$, o aludido paradoxo de Klein, implicando que mais
part\'{\i}culas s\~{a}o refletidas na barreira de potencial que aquelas
incidentes. Tem que ser assim porque, conforme vimos anteriormente, o
componente vetorial da barreira de potencial estimula a produ\c{c}\~{a}o de
antipart\'{\i}culas em $x=0$. Em virtude da conserva\c{c}\~{a}o da \textit{%
carga}\ h\'{a}, em verdade, a cria\c{c}\~{a}o de pares part\'{\i}%
cula-antipart\'{\i}cula e, como o potencial vetorial em $x>0$ \'{e}
repulsivo para part\'{\i}culas, elas ser\~{a}o necessariamente refletidas. N%
\~{a}o apenas a\ \textit{carga}\ \'{e} conservada. Visto que os pares
produzidos em $x=0$ t\^{e}m energias de sinais contr\'{a}rios, conclui-se
que a energia tamb\'{e}m \'{e} uma quantidade conservada no processo de cria%
\c{c}\~{a}o de pares.

\subsection{O limiar para a produ\c{c}\~{a}o de pares}

Da discuss\~{a}o relacionada com as classes \textbf{B} e \textbf{C},
observa-se que o limiar para a produ\c{c}\~{a}o de pares \'{e} dado por $%
V_{c}$. Donde torna-se evidente que o acoplamento escalar resulta no aumento
da energia m\'{\i}nima necess\'{a}ria para a cria\c{c}\~{a}o de pares part%
\'{\i}cula-antipart\'{\i}cula. O valor m\'{\i}nimo do limiar ($%
V_{0}=2mc^{2}) $ ocorre quando o acoplamento \'{e} puramente vetorial ($%
g_{t}=1$). A adi\c{c}\~{a}o de um contaminante escalar contribui para
aumentar o valor do limiar, o qual, surpreendentemente, torna-se infinito j%
\'{a} para uma mistura meio a meio de acoplamentos. Deste modo, a produ\c{c}%
\~{a}o de pares n\~{a}o \'{e} fact\'{\i}vel se o acoplamento vetorial n\~{a}%
o exceder o acoplamento escalar, ainda que o potencial $V_{0}$ seja
extremamente forte.

Pode-se interpretar a possibilidade de propaga\c{c}\~{a}o de antipart\'{\i}%
culas al\'{e}m da barreira de potencial como sendo de\-vi\-do ao fato que \
cada antipart\'{\i}cula est\'{a} sujeita a um potencial efetivo dado por $%
\left( g_{s}-g_{t}\right) V_{0}$, destarte se $g_{t}>1/2$ a antipart\'{\i}%
cula ter\'{a} uma energia dispon\'{\i}vel (energia de repouso mais energia
cin\'{e}tica) expressa por $\left( 2g_{t}-1\right) V_{0}-E$, donde se
conclui sobre a energia do limiar da produ\c{c}\~{a}o de pares. Pode-se
afirmar ainda que as part\'{\i}culas est\~{a}o sob a influ\^{e}ncia de um
potencial degrau ascendente de altura $V_{0}=\left( g_{s}+g_{t}\right) V_{0}$%
, e que as antipart\'{\i}culas est\~{a}o sujeitas a um potencial degrau
efetivo de altura $\left( g_{s}-g_{t}\right) V_{0}$, um degrau ascendente
(repulsivo) se $g_{t}<1/2$ e descendente (atrativo) se $g_{t}>1/2$.

\subsection{A penetra\c{c}\~{a}o na regi\~{a}o classicamente proibida}

Investigamos agora o efeito da onda evanescente em $x>0$, relacionado com a
classe \textbf{B}. Neste caso, o estado estacion\'{a}rio al\'{e}m da
barreira de potencial \'{e} descrito pela autofun\c{c}\~{a}o $\phi
=B_{+}\,e^{-|\kappa |x}$, de modo que a incerteza na posi\c{c}\~{a}o,
estimada como sendo o valor de $x$ que torna a densidade igual a $1/e$ de
seu valor em $x=0$, redunda em $\Delta x=1/(2|\kappa |)$, como acontece na
teoria qu\^{a}ntica n\~{a}o-relativ\'{\i}stica. Entretanto, contrariamente
\`{a} previs\~{a}o da teoria n\~{a}o-relativ\'{\i}stica, $\Delta x$
apresenta o valor m\'{\i}nimo%
\begin{equation}
\left( \Delta x\right) _{\min }=\frac{\hbar }{2\left( mc+g_{s}V_{0}/c\right)
}  \label{17}
\end{equation}%
quando $V_{0}$ torna-se igual a
\begin{equation}
V_{m}=\frac{E}{g_{t}}  \label{17a}
\end{equation}%
Por meio desta \'{u}ltima express\~{a}o vemos que $\left( \Delta x\right)
_{\min }=\lambda/2$ no caso de um potencial vetorial puro ($g_{s}=0$),
em harmonia com o pr\'{\i}ncipio da incerteza. Contudo, podemos concluir que
$\left( \Delta x\right) _{\min }<\lambda/2$ no caso de um potencial
vetorial contaminado com algum acoplamento escalar. \`{A} primeira vista
isto parece um resultado desatroso por violar o princ\'{\i}pio da incerteza
de Heisenberg. Liberta-se desta dana\c{c}\~{a}o considerando-se que o
componente escalar do potencial contribui para alterar a massa da part\'{\i}%
cula. Realmente, definindo a massa efetiva como $m_{\mathtt{ef}%
}=m+g_{s}V_{0}/c^{2}$ segue-se imediatamente que $\left( \Delta x\right)
_{\min }=\lambda_{\mathtt{ef}}/2$ \ e $\left( \Delta p\right) _{\max
}=m_{\mathtt{ef}}c$, onde o comprimento de onda Compton efetivo \'{e}
definido como $\lambda_{\mathtt{ef}}=\hbar /(m_{\mathtt{ef}}c)$.

\section{Conclus\~{a}o}

Exploramos a EKG em uma dimens\~{a}o espacial por motivos de simplicidade.
Consideramos potenciais externos com a mais geral estrutura de Lorentz e
mostramos que, se a intera\c{c}\~{a}o escalar \'{e} acoplada adequadamente,
a EKG independente do tempo reduz-se \`{a} equa\c{c}\~{a}o de Schr\"{o}%
dinger independente do tempo no limite n\~{a}o-relativ\'{\i}stico.

A an\'{a}lise do potencial degrau com uma mistura arbitr\'{a}ria de
acoplamentos vetorial e escalar mostrou-se muito prof\'{\i}cua. Tr\^{e}s
classes de solu\c{c}\~{o}es foram discernidas. Em todas essas tr\^{e}s
classes, o acoplamento escalar n\~{a}o desempenha papel expl\'{\i}cito na
determina\c{c}\~{a}o da velocidade de grupo, e nenhum papel na determina\c{c}%
\~{a}o da densidade e da corrente.

A mistura arbitr\'{a}ria de acoplamentos no potencial degrau desvelou a
i\-ne\-xe\-q\"{u}i\-bi\-li\-da\-de do me\-ca\-nis\-mo da produ\c{c}\~{a}o
espont\^{a}nea de pares no caso em que $g_{t}\leq 1/2$, tanto quanto o
aumento do limiar da energia de produ\c{c}\~{a}o no caso em que %
\mbox{$g_{t}>1/2$}. Outrossim, a presen\c{c}a de um acoplamento escalar
revelou a possibilidade de localizar part\'{\i}culas em regi\~{o}es do espa%
\c{c}o arbitrariamente pequenas. Com efeito, a presen\c{c}a de um
acoplamento escalar, por menor que seja, conduz a $\ \left( \Delta x\right)
_{\min }\rightarrow 0$ quando $V_{0}\rightarrow \infty $ sem que haja
qualquer chance para a produ\c{c}\~{a}o de pares na interface dos
potenciais. Isto dito tendo em vista que $V_{m}$, o potencial que minimiza a
incerteza na posi\c{c}\~{a}o, \'{e} sempre menor que $V_{c}$, o potencial do
limiar da produ\c{c}\~{a}o espont\^{a}nea de pares.

O limite n\~{a}o-relativ\'{\i}stico da EKG expresso pela Eq. (\ref{3b}), ou
seja a equa\c{c}\~{a}o de Schr\"{o}dinger com e\-ner\-gia de liga\c{c}\~{a}o
$E-mc^{2}$, n\~{a}o diferencia o acoplamento vetorial do acoplamento escalar
e pressup\~{o}e que $V_{0}<<mc^{2}$. Portanto, conclui-se seguramente que n%
\~{a}o h\'{a} produ\c{c}\~{a}o de pares ($\left( V_{c}\right) _{\min
}=2mc^{2}$) nem incerteza m\'{\i}nima na posi\c{c}\~{a}o ($\left(
V_{m}\right) _{\min }=mc^{2}$) no regime n\~{a}o-relativ\'{\i}stico da EKG,
como \'{e} esperado.

Naturalmente, os coeficientes de reflex\~{a}o e transmiss\~{a}o para a
classe de solu\c{c}\~{o}es que envolve a cria\c{c}\~{a}o de pares foram
determinados de maneira aproximada, por\-quan\-to descuidou-se da intera\c{c}%
\~{a}o \textit{interna} entre part\'{\i}culas e antipart\'{\i}culas.

Finalmente, ainda que haja intera\c{c}\~{o}es externas extremamente fortes,
a inviabilidade do me\-ca\-nis\-mo de produ\c{c}\~{a}o de pares no caso em
que o acoplamento escalar excede o acoplamento vetorial parece preservar a
interpreta\c{c}\~{a}o do mo\-de\-lo de part\'{\i}cula \'{u}nica da EKG.
Entretanto, quando as condi\c{c}\~{o}es favor\'{a}veis ao me\-ca\-nis\-mo de
produ\c{c}\~{a}o de pares entram em cena, j\'{a} n\~{a}o se pode mais
esperar que o formalismo de primeira quantiza\c{c}\~{a}o seja satisfat\'{o}%
rio. Ainda que tais condi\c{c}\~{o}es n\~{a}o se manifestem, resta perguntar
qual o papel dos estados associados com as antipart\'{\i}culas. Mesmo na aus%
\^{e}ncia de potenciais externos, qual o me\-ca\-nis\-mo que evita que haja
transi\c{c}\~{o}es entre o estados de energia positiva pertencentes ao %
\mbox{continuum }entre $+mc^{2}$ e $+\infty $, e os estados de energia
negativa pertencentes ao \mbox{continuum }entre $-mc^{2}$ e $-\infty $? Eis
aqui exemplos de perguntas que encontram respostas satisfat\'{o}rias somente
no formalismo da segunda quantiza\c{c}\~{a}o da teoria.

\bigskip

\bigskip

\bigskip

\bigskip

\noindent \textbf{Agradecimentos:}

\medskip

Os autores s\~{a}o gratos ao CNPq e \`{a} FAPESP pelo apoio financeiro.

\end{document}